\documentclass[epsfig,graphics,twocolumn,showpacs,floatfix,mathbbm,pra,superscriptaddress]{revtex4}
\usepackage{graphicx}

\begin{document}

\title[a]{Quantum Storage of a Photonic Polarization Qubit in a Solid}%
\pacs{03.67.Hk,42.50.Gy,42.50.Md}
\author{Mustafa G\"{u}ndo\u{g}an}
\email{mustafa.gundogan@icfo.es}
\affiliation{ICFO-Institut de Ci\`{e}ncies Fot\`{o}niques, Mediterranean Technology Park, 08860 Castelldefels (Barcelona), Spain}%

\author{Patrick M. Ledingham}
\affiliation{ICFO-Institut de Ci\`{e}ncies Fot\`{o}niques, Mediterranean Technology Park, 08860 Castelldefels (Barcelona), Spain}%

\author{Attaallah Almasi}
\altaffiliation{Present address: Department of Physics and Astronomy and LaserLaB, VU Amsterdam, Amsterdam, 1081HV, The Netherlands.}
\affiliation{ICFO-Institut de Ci\`{e}ncies Fot\`{o}niques, Mediterranean Technology Park, 08860 Castelldefels (Barcelona), Spain}%

\author{Matteo Cristiani}
\affiliation{ICFO-Institut de Ci\`{e}ncies Fot\`{o}niques, Mediterranean Technology Park, 08860 Castelldefels (Barcelona), Spain}%

\author{Hugues de Riedmatten}
\affiliation{ICFO-Institut de Ci\`{e}ncies Fot\`{o}niques, Mediterranean Technology Park, 08860 Castelldefels (Barcelona), Spain}%
\affiliation{ICREA-Instituci\'o Catalana de Recerca i Estudis Avan\c{c}ats, 08015 Barcelona, Spain}%


\begin{abstract}
We report on the quantum storage and retrieval of photonic
polarization quantum bits onto and out of a solid state storage
device. The qubits are implemented with weak coherent states at
the single photon level, and are stored for $500\,\mathrm{ns}$ in
a praseodymium doped crystal with a storage and retrieval
efficiency of $10\,\mathrm{\%}$, using the atomic frequency comb
scheme. We characterize the storage by using quantum state
tomography, and find that the average conditional fidelity of the
retrieved qubits exceeds $95\,\%$ for a mean photon number
$\mu=0.4$. This is significantly higher than a classical
benchmark, taking into account the Poissonian statistics and
finite memory efficiency, which proves that our device functions
as a quantum storage device for polarization qubits, even if
tested with weak coherent states. These results extend the storage
capabilities of solid state quantum memories to polarization
encoding, which is widely used in quantum information science.
\end{abstract}

\maketitle

The ability to transfer quantum information in a coherent,
efficient and reversible way from light to matter plays an
important role in quantum information science \cite{Hammerer2010}.
It enables the realization of photonic quantum memories (QM)
\cite{simon2010} which are required to render scalable elaborate
quantum protocols involving many probabilistic processes that have
to be combined. A prime example is the quantum repeater
\cite{Briegel1998,Duan2001,Sangouard2011}, where quantum
information can be distributed over very long distances. Other
applications include quantum networks \cite{Kimble2008}, linear
optics quantum computation \cite{Kok2007}, deterministic single
photon sources \cite{Matsukevich2006} and multiphoton quantum
state engineering.

Proof of principle experiments demonstrating photonic QMs have
been reported in different atomic systems such as cold
\cite{Chaneliere2005,Chou2005,Simon2007c,Choi2008,Zhao2009,Radnaev2010,Zhang2011}
and hot atomic gases \cite{Julsgaard2004,Eisaman2005, Reim2011,
Hosseini2011}, single atoms in cavities \cite{Specht2011} and
solid state systems \cite{Riedmatten2008,Hedges2010}. In recent
years, solid state atomic ensembles implemented with rare-earth
doped solids have emerged as a promising system to implement QMs.
They provide a large number of atoms with excellent coherence
properties naturally trapped into a solid state system. In
addition, they feature a static inhomogeneous broadening that can
be shaped at will, enabling new storage protocols with enhanced
storage properties (e.g. temporal multiplexing)
\cite{Kraus2006,Afzelius2009}. Finally, some of the rare-earth
doped crystals (e.g. praseodymium and europium doped crystals)
possess ground states with extremely long coherence times
\cite{Longdell2005} ($>$ seconds), which hold promise for
implementing long lived solid state quantum memories.

Recent progress towards solid state QMs include the storage of
weak coherent pulses at the single photon level
\cite{Riedmatten2008,Sabooni2010,Chaneliere2010,Lauritzen2010},
the quantum storage of coherent pulses with efficiency up to $70\,\%$ \cite{Hedges2010}, the spin state storage of bright coherent
pulses \cite{Longdell2005,Afzelius2010} and the storage of
multiple temporal modes in one crystal
\cite{Usmani2010,Bonarota2011}. Very recently, these capabilities
have been extended to the storage of nonclassical light generated
by spontaneous down conversion, leading to the entanglement
between one photon and one collective atomic excitation stored in
the crystal \cite{Clausen2011,Saglamyurek2011}, and entanglement
between two crystals \cite{Usmani2011}.

All previous experiments with solid state QMs have been so far
limited to the storage of multiple modes using the time degree of
freedom, e.g. time bin or energy time qubits. However, quantum
information is very often encoded in the polarization states of
photons, which provide an easy way to manipulate and analyze the
qubits. Extending the storage capabilities of solid state QMs to
polarization encoded qubits would thus bring much more flexibility
to this kind of interface. Unfortunately, storing coherently
polarization states is not straightforward in rare-earth doped
crystals. The main difficulty is that these crystals have in
general a strongly polarization dependent absorption. Storing
directly a polarization qubit in such a system would result in a
severely degraded fidelity of the retrieved qubits.

In this paper, we report on the  storage and retrieval of a
photonic polarization qubit into and out of a solid state quantum
storage device with high conditional fidelity. The photonic qubits
are implemented with weak coherent pulses of light, with a mean
photon number $\mu$ from $0.01$ to $36$. We measure the
conditional fidelity \cite{simon2010} of the storage and retrieval
process (i.e. assuming that a photon was re-emitted) and compare
it to classical benchmarks. With this procedure, we can show that
our crystal behaves as a quantum storage device, even if tested
with classical, weak coherent pulses. We overcome the difficulty
of anisotropic absorption by splitting the polarization components
of the qubit and storing them in two spatially separated ensembles
within the same crystal \cite{Matsukevich2004,Chou2007,Choi2008}.

\begin{figure}
   \centering
   \includegraphics[width=.45\textwidth]{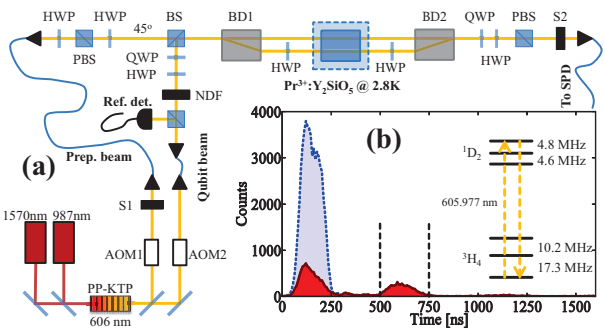}
   \caption{(Color online) (a) Experimental setup. The preparation beam and the qubit beam are
   derived from the same laser at $606\,\mathrm{nm}$.
    Both beams are amplitude and frequency modulated using an acousto-optic modulator (AOM).
    The polarization of the preparation beam
   is set to $45$ degrees, while the polarization of the qubit beam
   can be set arbitrarily with a half wave plate (HWP) and a
   quarter wave plate (QWP). The qubit beam is attenuated down to the single photon level using neutral density filters (NDF). The two beams are recombined at a
   beam splitter (BS) and sent to the storage device which is
   composed of a $\mathrm{Pr^{3+}:Y_2SiO_5}$ crystal cooled to $2.8\,\mathrm{K}$,
   two beam displacer (BD) and two HWP. The light released by the
   memory is then sent to a polarization analyzer composed
   of a HWP, a QWP and a polarization beam splitter (PBS), before
   being detected with a single photon detector (SPD). The two
   mechanical shutters (S1 and S2) are used to suppress optical noise from the preparation beam and to protect the
   SPD. (b) Storage and retrieval of a weak $|V\rangle$ qubit with
   duration $140\,\mathrm{ns}$ and $\mu=0.4$. The temporal histogram of the detection
   without (dotted line,empty pit) and with (solid line) AFC is shown.
   The dashed lines around the AFC echo show the detection
   window. Inset: Level scheme of $\mathrm{Pr^{3+}:Y_2SiO_5}$ with the dashed arrows showing the relevant transition for photon absorption and re-emission. }
   \label{setup}
   \end{figure}

Our memory is implemented using a $3\,\mathrm{mm}$ long
$\mathrm{Pr^{3+}:Y_2SiO_5}$ crystal ($0.05\,\%$). The relevant
atomic transition connects the $\mathrm{^3H_4}$ ground state to
the $\mathrm{^1D_2}$ excited state and has a wavelength of
$605.977\,\mathrm{nm}$. Each state has three hyperfine sublevels
as shown in Fig. \ref{setup}. The measured maximal optical depth
at the center of the $5\,\mathrm{GHz}$ inhomogeneous line is
$\mathrm{OD}=7$. We use the Atomic Frequency Comb (AFC) scheme to
store and retrieve the qubits \cite{Afzelius2009}. This requires
to shape the inhomogeneous absorption profile into a series of
periodic and narrow absorbing peaks, placed in a wide transparency
window. This creates a frequency grating and when a photon is
absorbed by the comb, it will be diffracted in time and re-emitted
after a pre-determined time $t_S=1/\Delta$, where $\Delta$ is the
spacing between absorption peaks.

Our experimental apparatus is described in Fig. \ref{setup}. The
laser source to generate light at $606\,\mathrm{nm}$ is based on
sum frequency generation (SFG) in a PP-KTP waveguide (AdVR corp)
from two amplified laser diodes at $1570\,\mathrm{nm}$ (Toptica,
DL 100 and Keopsys fiber amplifier) and $987\,\mathrm{nm}
$(Toptica, TA pro). With input power of $370\,\mathrm{mW}$ and
$750\,\mathrm{mW}$ for the $987\,\mathrm{nm}$ laser and
$1570\,\mathrm{nm}$ laser, respectively, we achieve an output
power of $90\,\mathrm{mW}$ at $606\,\mathrm{nm}$. Taking into
account the $30\,\%$ coupling efficiency of both beams into the
waveguide, we obtain a SFG efficiency of $\sim350\, \mathrm{\%
W^{-1}}$. The beam is then split in two parts, one which will be
used for the memory preparation (preparation beam) and one to
prepare the weak pulses to be stored (qubit beam). In each path,
the amplitude of the light is modulated with an acousto-optic
modulator (AOM) in a double pass configuration, in order to create
the required sequence of pulses for the preparation of the memory
and of the polarization qubits. The radio-frequency signals used
to drive the AOMs are generated by an arbitrary waveform generator
($500\,\mathrm{Msample/s}$, $200\,\mathrm{MHz}$, $\,\mathrm{1 GB}$
internal memory, PXIe module and ProcessFlow software from
Signadyne). After the AOMs, both beams are coupled to a
polarization maintaining optical fiber and sent to another optical
table where the cryostat is located.

The crystal is cooled down to $2.8\,\mathrm{K}$ in a cryofree
cooler (Oxford Instruments V14). After the fibers, the preparation
beam is collimated to a diameter of around $600\,\mathrm{\mu m}$
with a telescope and sent to the storage device. Right before the
cryostat, a beam displacer (BD1) splits the two polarization
components of the incoming light onto two parallel spatial modes
separated by $2.7\,\mathrm{mm}$, co-propagating through the
crystal. To ensure equal power in both spatial modes, the
polarization of the preparation beam is set to $45$ degrees. After
BD1, the polarization of the horizontal beam (lower beam in
Fig.\ref{setup}) is rotated by $90$ degrees using a HWP such that
both beams enter the crystal with the same polarization which is
parallel to the D2 axis of the crystal, which maximizes the
absorption. The qubit beam is strongly attenuated by a set of
fixed and variable neutral density filters (NDF), and $\mu$ before
BD1 was varied from $0.01$ to $36$. After the NDF, arbitrary
polarization qubits are prepared, using a quarter (QWP) and a half
wave plate (HWP). The qubits are then overlapped to the
preparation beam at a beam splitter (BS). After the cryostat, we
rotate back the polarization of the lower beam and the two spatial
modes are combined again at a second beam displacer (BD2). The two
path between BD1 and BD2 form an interferometer with very high
passive stability \cite{Chou2007,Choi2008}.

After the interferometer, the transmitted and retrieved light
enters the polarization analysis stage, composed of a QWP, a HWP
and a polarization beam splitter (PBS), which allow us to measure
the polarization in any basis. The transmitted beam at the PBS is
coupled in a multimode fiber and sent to a Silicon avalanche
photodiode Single Photon detector (SPD, model Count, Laser
Components). The electronic signal from the SPD is finally sent to
a time stamping card (PXIe card from Signadyne) in order to record
the arrival time histogram. The mean photon number $\mu$ is
determined by measuring the detection probability per pulse
$p_{det}$ when no atoms are present (i.e. with the laser $60\,\mathrm{GHz}$
off resonance), and backpropagating before BD1 taking into account
the detection efficiency ($\eta_D=50\,\%$) and the transmission
from before BD1 to the detector ($\eta_t=40\,\%$). The
preparation beam and the qubits are sent sequentially towards the
crystal. The total experimental sequence lasts $3$ seconds, during
which the preparation lasts $1\,\mathrm{s}$. During the next $2$ seconds $10^5$
weak pulses are prepared, stored and retrieved at a rate of $50\,\mathrm{kHz}$.

In order to create the AFC, we first create a wide transparency
window within the $5\,\mathrm{GHz}$ inhomogeneous profile. This is
achieved by sending a series of pulses of duration
$1.1\,\mathrm{ms}$, during which the frequency of the light is
swept linearly over a range of $12\,\mathrm{MHz}$. The AFC is then
created using the burn back procedure introduced in Ref.
\cite{Nilsson2004}, i.e. by illuminating the sample with short
pulses of duration $2\,\mathrm{ms}$, while shifting the frequency
of the light by $-27\,\mathrm{MHz}$ with respect to the center of
the pit. Four burn back pulses are sent with different frequencies
separated by the comb spacing, leading to a 4-tooth comb.

\begin{figure}
   \centering
   \includegraphics[width=.4\textwidth]{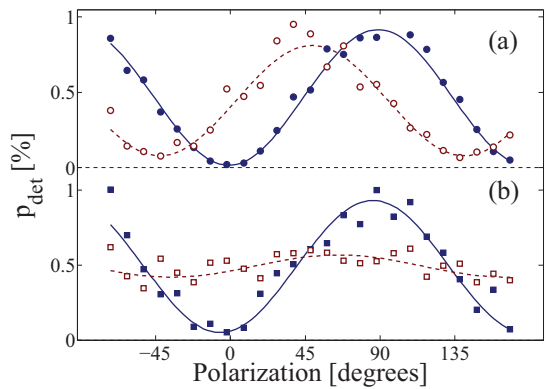}
   \caption{(Color online) (a) Measured detection probability ($p_{det}$)
   as a function of the polarizer angle, for $|V\rangle$ (filled circles) and $|D\rangle $
   (open circles) input polarization qubits, with $\mu=0.4$. The fitted raw visibilities are $(97 \pm 0.5)\,\%$ and $(83\pm 2)\,\%$, respectively.
   (b) $p_{det}$ as a function of polarization angle for $|R\rangle$
   polarization qubit input, with (filled squares, $(89 \pm 2)\,\%$) and without (open squares, $(15 \pm 3)\,\%$) QWP inserted before the polarizer.}
   \label{fringe}
   \end{figure}

   \begin{figure}
   \centering
   \includegraphics[width=.42\textwidth]{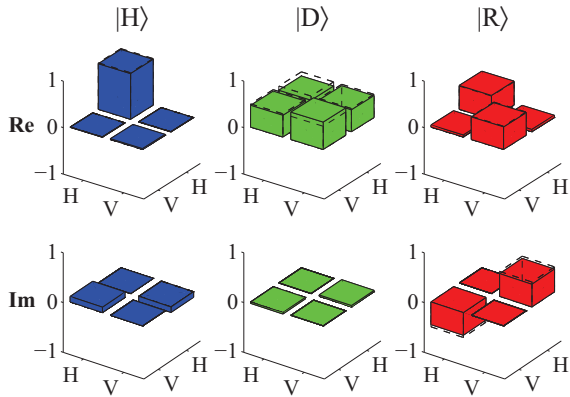}
   \caption{(Color online) Quantum state tomography. Reconstructed density matrices of the retrieved qubits for $|H\rangle$, $|D\rangle$ and $|R\rangle$
   input qubits, with $\mu =0.4$ .
   No background has been subtracted. }
   \label{tomog}
   \end{figure}

\begin{table}[htdp]
\begin{center}
\begin{tabular}{|c|c|c|c|}
\hline Input State & Fidelity & Input State & Fidelity \\ \hline
 $|H\rangle$& 0.982 $\pm$ 0.003& $|V\rangle$ & 0.983 $\pm$ 0.002\\
 $|D\rangle$ & 0.968 $\pm$ 0.005& $|A\rangle$ & 0.938 $\pm$ 0.009\\
 $|R\rangle$ & 0.954 $\pm$ 0.007 &  $|L\rangle$ & 0.926 $\pm$ 0.01\\
 \hline
  \end{tabular}
\end{center}

\caption{Raw conditional fidelities for 6 different polarization
input states. For this measurement, $\eta_M$ varied between 8 $\%$
and 10 $\%$. The errors have been obtained using Monte Carlo
simulation taking into account the statistical uncertainty of
photon counts and a technical error reflecting slow drifts in our
systems, and estimated from the residuals from the fit of
Fig.\ref{fringe} and similar curves. } \label{tab:fidel}
\end{table}
 As a first experiment, we
verified that a complete set of qubit distributed over the
Poincar\'{e} sphere could be stored and retrieved in the AFC. We
set the storage time to $500\,\mathrm{ns}$ and we used the
following input states: $|H\rangle$,$|V\rangle$, $ |D\rangle =
\frac{1}{\sqrt{2}}(|H\rangle+|V\rangle)$, $|A\rangle =
\frac{1}{\sqrt{2}}(|H\rangle-|V\rangle)$, $|R\rangle =
\frac{1}{\sqrt{2}}(|H\rangle+i|V\rangle)$ and $|L\rangle =
\frac{1}{\sqrt{2}}(|H\rangle-i|V\rangle)$. Fig.\ref{setup} (b)
shows the experimental storage of a $|V\rangle$ qubit encoded onto
a pulse of duration $140\,\mathrm{ns}$ (FWHM) and with $\mu =
0.4$. Similar curves are obtained for the other states, and the
average storage and retrieval efficiency is $\eta_M= (10.6 \pm
2.3)\,\%$. In order to test that the coherence between the $
|H\rangle$ and $ |V\rangle$ components of the qubits is preserved
during the storage and retrieval, we then recorded the number of
counts in the retrieved pulses when rotating the detection
polarization basis using a HWP, for various input states. In Fig.
\ref{fringe} (a), the curves obtained for $ |V\rangle$ and $
|D\rangle$ are shown. We observe interference fringes, with a raw
fitted visibility of $(97\pm0.5)\,\%$ for the $ |V\rangle$ qubit,
and $(83\pm 2)\,\%$ for the $|D\rangle$ qubit. In the case of a
perfect circular $|R\rangle$ qubit, we should observe no
dependance on the HWP angle when no QWP is inserted. The
interference should be restored however, when a QWP is inserted
before the HWP, which turns circular polarization into a linear
one. The results are shown in Fig. \ref{fringe} (b). We indeed
observe a strongly reduced visibility without the HWP ($(15\pm
3)\,\%$), while a fringe with a high visibility of $(89\pm 2)\,\%$
is obtained with the QWP. The residual visibility without the QWP
may be due to a non perfect preparation of the $|R\rangle$ state.
The non perfect visibility for the $|D\rangle$ and $|R\rangle$ is
mostly due to small phase fluctuations engendered by mechanical
vibrations from the cryostat. Indeed, we observe similar
visibilities for bright pulses out of resonance with the atomic
transition. These results show that the phase between the two
polarizations components is almost perfectly preserved in the
storage and retrieval process, and is preserved to a high degree
in our combined interferometer and memory setup, for various
polarization input states.

\begin{figure}
   \centering
   \includegraphics[width=.45\textwidth]{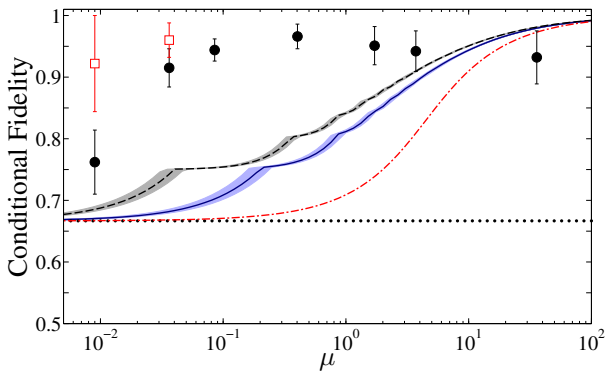}
   \caption{(Color online) Average fidelity measured as a function of the mean number of photon per pulse $\mu$.
The fidelity is measured by quantum state tomography and is the
average of  $3$ input states $ |V\rangle$, $ |D\rangle$, and $
|R\rangle$.  Filled circles are experimental data without any
background subtraction. Empty squares correspond to dark count
subtracted fidelities. The error takes into account statistical
uncertainty of photon counts, technical errors and standard
deviation of fidelities for the 3 polarizations. The various lines
correspond to classical thresholds for different situations. The
horizontal line is the limit of $2/3$ for single qubits (N=1). The
dashed-dotted line corresponds to Eq.\ref{fidelity} where the
Poissonian distribution of photon number is taken into account.
Finally, the two other lines correspond to the cases where the
finite storage efficiency is taken into account. The solid line
corresponds to $\eta=(10\pm 2)\,\%$. Measured $\eta_M$ are between
$8$ and $10\,\%$ for all points. The dashed line corresponds to
$\eta=\eta_M\eta_t\eta_D=2\,\%$. For $\mu> 1$, an additional ND
filter is placed before the detector to avoid saturation effects.}
   \label{fidelity_vs_N}
   \end{figure}

In order to better characterize the quality of the storage
process, we reconstruct the density matrix of the retrieved qubits
using quantum state tomography \cite{James2001}, for the complete
set of qubits described above, with $\mu = 0.4$. The reconstructed
output density matrices $\rho_{out}$ for $|H\rangle$, $|D\rangle$
and $|R\rangle $ are shown in Fig \ref{tomog}. From the matrices
$\rho_{out}$, we can then estimate the conditional fidelity of the
output states with respect to the target state
$F^c_{|\psi\rangle}=\langle\psi|\rho_{out}|\psi\rangle$. The
values for the complete set of inputs are listed in Table 1. We
find a mean fidelity of $F^c_{mean}=(96\pm2)\,\%$.  We emphasize
that this value is a lower bound for the conditional fidelity,
since it is calculated with respect to a target state and also
takes into account imperfections in the preparation of the qubits.

Finally, in order to assess the quantum nature of the storage, we
determine the average fidelity  as a function of $\mu$, and
compare it with the best obtainable fidelity using a purely
classical method consisting of measuring the state and storing the
result in a classical memory. It has been shown that for a state
containing $N$ qubits, the best classical strategy leads to a
fidelity of $F_c=(N+1)/(N+2)$ \cite{Massar1995}, leading to the
well known fidelity of $2/3$ for $N=1$. If the qubit is encoded in
a weak coherent state, as it is the case in our experiment, one
has to take into account the Poissonian statistics of the number
of photons \cite{Specht2011}, and the classical fidelity is given
by:
\begin{equation}\label{fidelity}
   F_{class} (\mu)=\sum_{N\geq1} \frac{N+1}{N+2}
   \times\frac{P(\mu,N)}{1-P(\mu,0)},
\end{equation}

where $P(\mu,N)=e^{-\mu}\mu^N/N!$. This is valid for the case of a
memory with unity efficiency. If $\eta_M <1$, the classical memory
could use a more elaborate strategy to take advantage of finite
efficiency in order to gain more information about the input
quantum state \cite{Specht2011}. For example the classical memory
could give an output only when the number of photons per pulse is
high, and hence estimate with better fidelity the quantum state
(See appendix \ref{appdx}). The different curves
corresponding to the discussed cases are plotted in Fig.
\ref{fidelity_vs_N} as a function of $\mu$. The points correspond
to experimental data. Measured $F^c_{mean}$ are significantly
higher than the classical fidelity, for most of the photon numbers
tested. This proves that our device performs as a quantum storage
device for polarization qubits, even if tested with weak coherent
states. We observe that the measured raw fidelity decreases for
$\mu < 0.1$. This is mainly due to the dark count of the SPD, as
high fidelities can be recovered by subtraction of this background
(open squares). We also observe that when $\mu$ becomes too large
($\mu \geq 3.5$ in our case), the measured fidelity is not
sufficient to be in the quantum regime. This confirms that very
low photon numbers are required to test the quantum character of
QMs with weak coherent states. To our knowledge, it is the first
time that an ensemble based memory has been characterized using
this criteria.

The storage time in our experiment is limited by the minimal
achievable width of the AFC peaks ($<$600 kHz), which is in turn
fully limited by the linewidth of our unstabilized laser. Peaks as
narrow as $30\,\mathrm{kHz}$ have been created in
$\mathrm{Pr^{3+}:Y_2SiO_5}$ using a frequency stabilized laser
\cite{Hetet2008a}, which should allow a storage time in the
excited state of about 10 $\mathrm{\mu s}$. This should also allow
the storage of multiple polarization qubits in the time domain. In
order to increase the storage time and to achieve on demand read
out with an AFC, the optical excitations should be transferred to
long lived spin excitation as demonstrated for bright pulses in
\cite{Afzelius2010}.

We have demonstrated the quantum storage and retrieval of
polarization qubits implemented with weak coherent pulses at the
single photon level, in a solid state storage device. The
conditional fidelity of the storage and retrieval is $>95\,\%$,
significantly exceeding the classical benchmark calculated for
weak coherent pulses and finite memory efficiency. We thus show
that solid state QMs are compatible with photonic polarization
qubits, which are widely used in quantum information science. This
significantly extends the storage capabilities of these types of
memories. By combining the time and polarization degrees of
freedom one could readily double the number of modes that can be
stored in the memory and create quantum registers for polarization
qubits. Using these resources, it may also be possible to design a
quantum memory for complex light states such as hyperentangled
states.

We note that related results have been obtained by two other groups \cite{Saglamyurek2011a, Clausen2012}.

We thank Stephan Ritter and Antonio Acin for interesting
discussions regarding the classical benchmark, and the company
Signadyne for technical support. Financial support by the
CHIST-ERA European project QScale and by the ERC Starting grant
QuLIMA is acknowledged.

\appendix
\section{Conditional Fidelity using Classical State Estimation for Weak Coherent States}
\label{appdx}

In this section, we give more details on the calculation of the
classical benchmark presented in Fig. $4$ of the main manuscript.
The general idea is to estimate what is the best efficiency that
can be obtained using a classical method. In particular, we
consider a measure and prepare strategy where the user performs a
classical state estimation on the input qubit, stores the result
in a classical memory and prepares a new qubit according to the
result obtained. The maximum achievable classical fidelity for a
state with a fixed photon number $N$ is known to be
\cite{Massar1995}
\[
F= \frac{N+1}{N+2}.
\]
If one tests a memory with a true single photon input then the
classical bound is ${2}/{3}$. If one uses weak coherent states
then one has to consider the finite probability of having more
than one photon. A pulse of light with mean photon number ${\mu}$
has a probability distribution of a Poissonian, given by
\[
P({\mu},N) = \mathrm{e}^{-{\mu}} \,\frac{{\mu}^N}{N}.
\]
Then, the maximum achievable fidelity becomes a weighted sum over $N$ of
the fidelity for a given $N$ where the weight is given by the
Poissonian statistics of the input \cite{Specht2011}. We state this below as
\begin{equation}
F = \sum_{N\geq 1}^{\infty} \; \frac{N+1}{N+2} \; \frac{P({\mu},N)}{1 - P({\mu},0)}.
\label{eq:F100}
\end{equation}

This formula is valid if one assumes that the quantum
memory has a storage and retrieval efficiency $\eta_M = 1$. If
$\eta_M <  1$, a more sophisticated classical strategy could
simulate non-unit efficiency by only giving a result for high
photon number $N$, resulting in a higher achievable fidelity, as
suggested in \cite{Specht2011}. We consider that the classical
measure and prepare strategy has an efficiency of 1, but we define
the effective classical efficiency $\eta_C$ as the probability
that the classical device gives an output qubit, if it has
received at least one photon as input:

\[
\eta_C = \frac{P_{\mathrm{out}}}{1 - P(\mu, 0)}.
\]

where
\[
P_{\mathrm{out}} = \sum_{ N\geq N_{\mathrm{min}}+1} P(\mu,
N).
\]

Note that the photon number statistics of the output qubit is not
relevant in our case, since we use non photon number resolving
detectors. The classical memory gives a result for some threshold
photon number $N_{\mathrm{min}}+1$ and above, and no result for
$N$ lower than this. It is important to note that for the above
there exists only certain $\eta_{\mathrm{C}}$ for a given mean
photon number $\mu$. Hence, not all quantum efficiencies can be
simulated.  A more general form of $P_{\mathrm{out}}$ which allows
for arbitrary number and hence efficiency is the following
\[
P_{\mathrm{out}} =\gamma \; + \sum_{N \geq N_{\mathrm{min}}+1}
P(\mu, N).
\]
where now the memory gives a result for $N_{\mathrm{min}}$ with a probability of $\gamma$, with the condition $\gamma \leq P(\mu,N_{\mathrm{min}})$.

\begin{figure}[h!]
   \centering
   \includegraphics[width=0.45\textwidth]{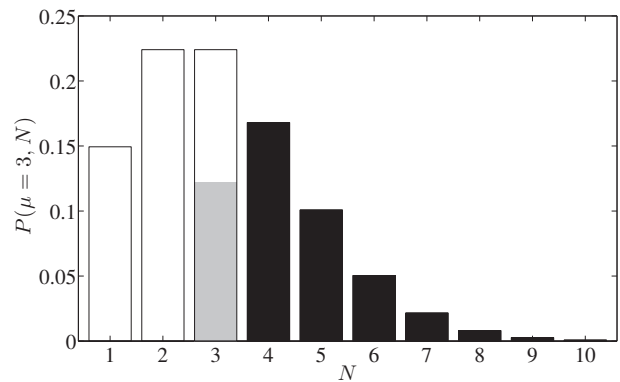}
   \caption{An example of obtaining $N_{\mathrm{min}}$. Plotted above is the probability distribution for a coherent state of $\mu = 3$. The target efficiency here is $50\%$. For this target efficiency $N_{\text{min}} = 3$, found using Equation \ref{eq:Nmin}. The grey shaded part of bar $N = 3$ plus the black bars $N \geq 4$ equate to the $50\%$ target efficiency.}
   \label{fig:Nmin}
\end{figure}

\begin{figure}[t!]
   \centering
   \includegraphics[width=0.45\textwidth]{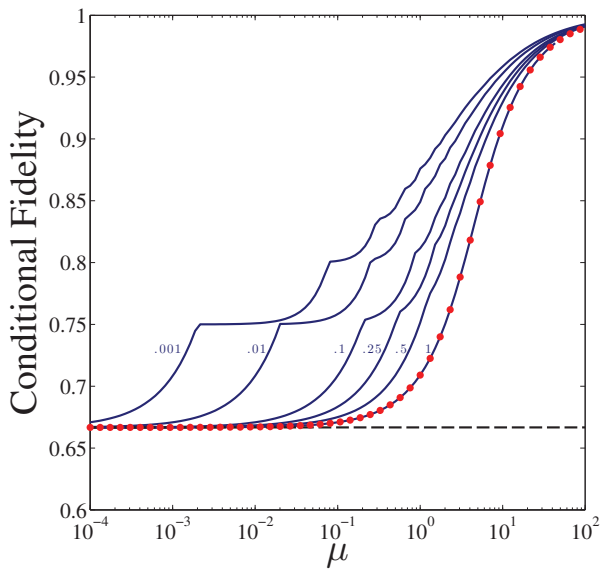}
   \caption{Maximum achievable fidelity using a classical memory as a function of mean photon number $\mu$ for different memory efficiencies $\eta$ (Eq. \ref{eq:eta}). The cases shown are $\eta~=~\{0.001, 0.01, 0.1, 0.25, 0.5, 1\}$. The solid lines are calculated using Equation \ref{eq:F}, the dots are calculated using Equation \ref{eq:F100}. It is seen that for $\eta = 1$, Equation \ref{eq:F} reduces to Equation \ref{eq:F100}. The dashed line is for the single photon case $N = 1$.}
   \label{fig:F}
\end{figure}

The efficiency is then
\begin{equation}
\eta_C =  \frac{\gamma \; + \sum_{N \geq N_{\mathrm{min}}+1} \,
P(\mu, N)}{1 -  P(\mu, 0)}. \label{eq:eta}
\end{equation}
We now assume that $\eta_M=\eta_C$ and $N_{\mathrm{min}}$ is
obtained as follows
\begin{equation}
N_{\min} = \mathrm{min }\,i :  \sum_{N \geq i +1}  P(\mu, N) \leq (1
-  P(\mu, 0))\,\eta_M.\label{eq:Nmin}
\end{equation}
Figure \ref{fig:Nmin} shows a graphical representation of
obtaining $N_{\text{min}}$.
The maximum achievable classical fidelity using the above
described strategy is then
\begin{equation}
F_{\mathrm{class}} = \frac{\left(\displaystyle\frac{N_{\mathrm{min}}
+1}{N_{\mathrm{min}} +2}\right) \gamma + \displaystyle \sum_{N\geq
N_{\mathrm{min}} +1} \displaystyle\frac{N+1}{N+2} P(\mu,N)
}{\gamma + \displaystyle\sum_{N\geq N_{\mathrm{min}} + 1} P(\mu,
N)} \label{eq:F},
\end{equation}
where $\gamma$ and  $N_{\mathrm{min}}$ are obtained from Equations (\ref{eq:eta}, \ref{eq:Nmin}). Figure \ref{fig:F}
shows the fidelity as a function of mean photon number $\mu$ for various efficiencies $\eta$.

Note that the classical memory could also in principle take
advantage of the optical loss and the finite detection
efficiencies in the experiment to increase the maximal classical
fidelity. In that case, we would have $\eta_C=\eta_M\eta_t\eta_D$,
where $\eta_t$ is the optical transmission from the quantum memory
to the detector and $\eta_D$ is the detection efficiency of the
SPD. For our experiment, we have $\eta_M=0.1$, $\eta_t=0.4$ and
$\eta_D=0.5$ such that in that case $\eta_C=0.02$.

\bibliographystyle{prsty}

\end{document}